# Stable Thrust on a Finite-sized Magnet above a Meissner Superconducting Torus

Jose-Luis Perez-Diaz1, Efren Diez-Jimenez2, Ignacio Valiente-Blanco2 and Javier Herrero-de-Vicente2
*1 Instituto Pedro Juan de Lastanosa. Universidad Carlos III de Madrid. Butarque, 15. E28911 Leganes. Spain.*
*2 Departamento de Ingeniería Mecánica. Universidad Carlos III de Madrid. Butarque, 15. E28911 Leganes. Spain.*

Forces and torques exerted by a superconducting torus on a permanent magnet have been mapped. It is demonstrated that stable orbits exist. Moreover, provided that the magnet remains in any of these orbits, the first critical field in the superconductor is never overpassed and the superconductor remains in the Meissner state. The consequent absence of hysteresis makes these kinds of device perfect candidates for non-frictional bearings or gyroscopes.

## I. INTRODUCTION

Repulsion forces between a Permanent Magnet and a Superconductor have frequently been proposed as candidates for producing bearings, gyroscopes and other kinds of mechanism. The absence of friction is by far their most remarkable and promising property.

However, so far the majority of experimental mechanical devices of this kind have utilized superconductors, but in the mixed and not the Meissner state. In such cases the quantized penetration of the magnetic flux easily provides the stability desired together with undesired friction when attempting to change the flux.

In contrast, the magnetization of a superconductor in the Meissner state is non-hysteretic, and consequently neither are the levitation forces. A pure Meissner state does not produce any friction at all. How to design stable levitating mechanisms while keeping the superconductor in the Meissner state is therefore an open challenge with multiple potential uses.

The levitation of a magnet above a superconductor in the Meissner state has been widely studied for different geometries: infinite planes, cylinders and spheres[1],[2],[3]. The method of the images can be used to provide analytical solutions for these highly symmetric cases.

For more complex shapes the analytical method of the images is not very practical, but a local model suitable for a numerical approach is available[4]. Following this approach both the force and torque between the magnet and the superconductor can be easily computed, whatever their shape[5],[6],[7]. In fact, in previous work one of the authors demonstrated that stability in one degree of freedom for a magnet levitating above a superconducting cylinder could be achieved[6].

In previous work, the levitation of a permanent magnet above a Meissner superconducting torus has been analyzed for a limited number of points in the axis using a finite elements approach according to the above-mentioned model[8],[9]. It was demonstrated that a phenomenon termed the "flip effect" appears for a magnet along the axis of the torus.

In the present work a large region is mapped in order to determine all the stable points and orbits.

## II. MODEL

The above-mentioned model was previously used by some of the authors to compute the force and torques for the case of a full superconducting cylinder and that of a hollow superconducting cylinder[10]. The method has been shown to be in a good agreement with the experimental results[4]. The model is valid for a complete Meissner state, i.e. the applied magnetic field is lower than the first critical field. It is purely based on London and Maxwell's equations.

According to this model the elementary force exerted by an external magnetic field per unit surface on the superconductor may be written, using the MKS system, as:

$$\frac{d\vec{F}}{dS} = 2\mu_0 (\vec{n}_s \times \vec{H}^{ap}) \times \vec{H}^{ap} \quad (1)$$

where $\mu_0$ is the magnetic permeability of a vacuum (or air), $\vec{n}_s$ is the unit vector orthogonal to the surface and $\vec{H}^{ap}$ is the magnetic field applied.

Moreover, for any shape of the superconductor, the torque applied to the superconductor may be written as:

$$\vec{M} = \iint_\Omega \vec{r} \times (2\mu_0 (\vec{n}_s \times \vec{H}^{ap}) \times \vec{H}^{ap}) dS \quad (2)$$

In order to calculate the moment applied on the magnet, $\vec{r}$ must be the position vector between the differential surface element and the center of mass of the permanent magnet.

Due to its local character (1) and (2) can be implemented in a finite element method program (FEA).

For the purpose of this paper a superconducting torus of 7.5 mm minor radius and 15 mm major radius has been modeled. A 3 mm diameter and 6 mm in height axially magnetized cylinder has been used as a permanent magnet.

This is a commercially-available variety of permanent magnet. A remanence of 1.16 T and a coercivity of 850

kA/m will be used – corresponding to linear $Nd_2Fe_{14}B$. The weight of such a magnet is $3.1 \cdot 10^{-3}$ N.

A convenient material for the torus could be the well-known High Tc Type II Superconductor $YBa_2Cu_3O_{7-x}$. For practical purposes we can assume liquid nitrogen is used to cool it at 77 K. This determines the maximum surface current induced on the superconductor compatible with a Meissner state[11].

We will use a coordinate system such that the z-axis is aligned with the axis of symmetry of the torus and X and Y are in its equatorial plane, as shown in Figure 1. The finite element model used is fully 3D, but the magnet in the positions of equilibrium is always in the poloidal xz-plane. Therefore, coordinates x and z will be enough to describe the position of the center of the magnet, as shown in Figure 1.

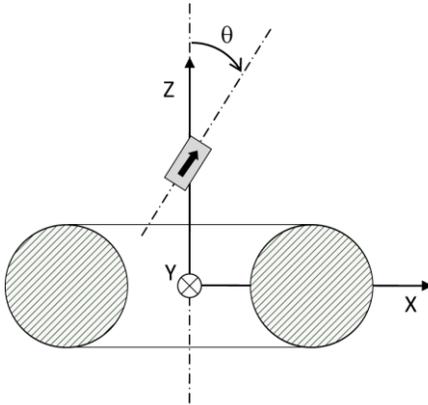

**Figure 1 – Schematic representation of the permanent magnet above the superconducting torus and coordinate system.**

Due to the symmetry of the torus, only the upper right part (x>0 and z>0) will be mapped.

Three cases will be considered for the mapping of the force and torque. The first two cases will be a magnet rotationally constricted to a fixed orientation: with the first case being that of a vertically-oriented magnet (parallel to the z-axis). The second case will be that of a horizontally-oriented magnet (parallel to the x-axis). If there are equilibrium points for a magnet levitating above the torus, there must also be equilibrium points in one of these two cases.

The third case will be a magnet that is free to rotate. In this case the angle of equilibrium for any position will also be calculated.

In all cases the region for which the vertical component of the levitating force $F_z$ equals the weight of the magnet will also be calculated. This will constitute the effective surface of levitation and will be called the "weight surface".

The region of the positions of the magnet for which the maximum current is overpassed on the torus ("Meissner surface") is also computed. In reality the model is only valid in cases where this current is not overpassed.

Since the force and torque are quadratic with respect to the magnetization of the magnet, the results can be easily scaled for magnets made with different materials. Therefore mapping will be offered even for those points for which the current on the torus clearly overpasses the Meissner maximum for the above-mentioned magnet made of $Nd_2Fe_{14}B$.

## III. VERTICAL MAGNET ($\theta = 0°$)

The components of the force $F_x$ and $F_z$, as well as component $M_y$ of the torque experienced by the magnet, are mapped in Figure 2. The Meissner surface is shown as a dotted line, while the weight surface is shown as a solid line.

$F_z$ is always positive, but it is null at the equatorial plane.

The upper weight surface is stable, while the lower and outer part of the weight surface are unstable. The stability of a weight surface occurs when $F_z$ decreases for an increasing z.

$F_x$ presents very different behavior in the inner and the outer part of the torus. It is positive in the outer part (x>15 mm), tending to expel the magnet away from the torus. It is negative in the inner part, driving the magnet towards the z-axis. The change from positive to negative is very sudden, and we can say that there is a discontinuity – that appears as a white line in Figure 2.

The torque component, $M_y$, also presents a discontinuity. It is positive in the inner region, but negative in the outer region of the space. It is only null along the z-axis. It is stable from z=0 to z=11.30 mm, but unstable for z above 11.30 mm.

As the weight surface (solid line) cuts the z-axis at z = 9.25 mm, this point will be a point of stable equilibrium for the permanent magnet levitating above the superconducting torus. As this point is above the Meissner surface (dotted line) the model is valid and the whole torus will be in the Meissner state when the magnet is at the point of equilibrium. This demonstrates that a completely stable levitation can be achieved with a torus made of $YBa_2Cu_3O_{7-x}$ at 77 K in a pure Meissner superconducting state.

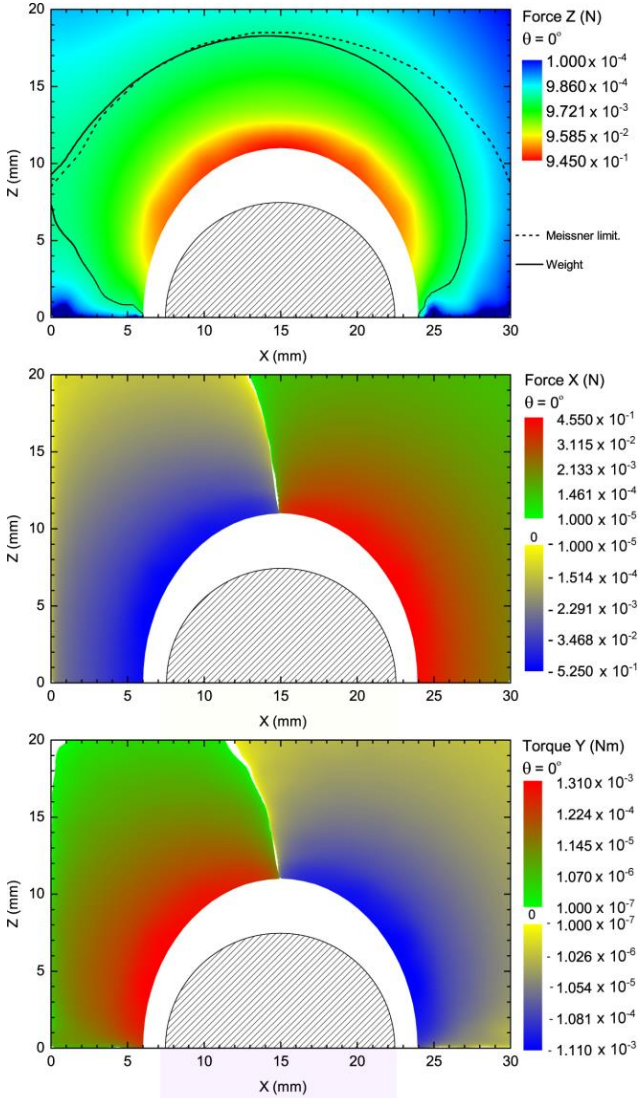

**Figure 2** – Vertical and lateral forces ($F_z$ and $F_x$) and torque ($M_y$) on a vertically-oriented magnet ($\theta = 0°$) above a superconducting torus.

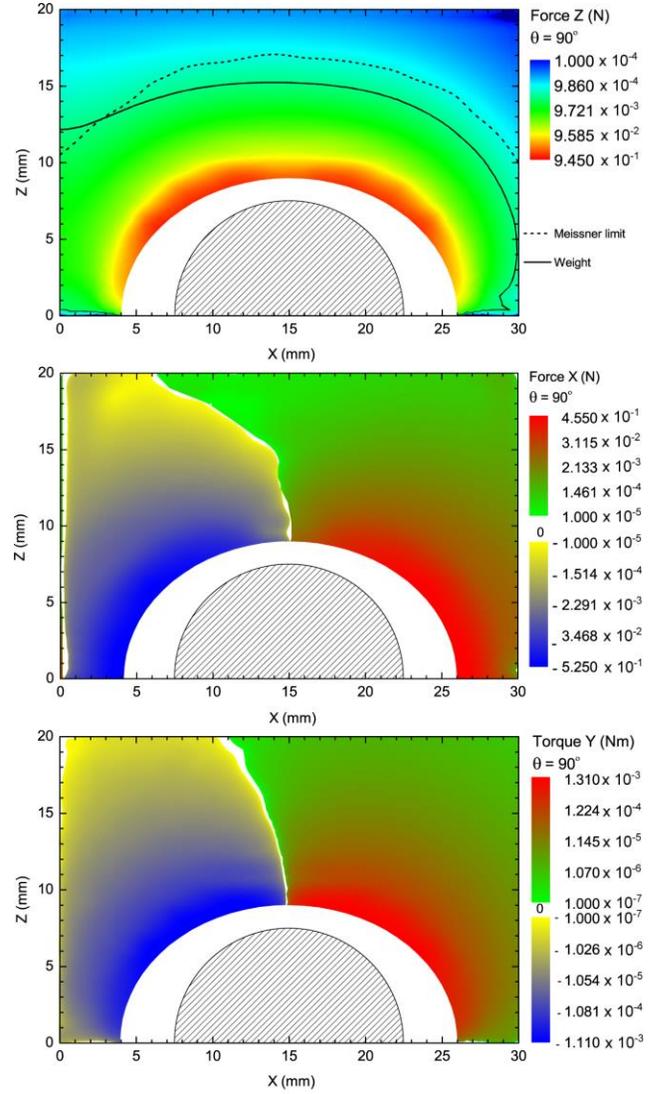

**Figure 3** – Vertical and lateral forces ($F_z$ and $F_x$) and torque ($M_y$) on a horizontally-oriented magnet ($\theta = 90°$) above a superconducting torus.

## IV. HORIZONTAL MAGNET ($\theta = 90°$)

The vertical $F_z$ and lateral $F_x$ forces and torque $M_y$ experienced by a horizontal magnet are mapped in Figure 3. The weight and Meissner surfaces are also shown as solid and dashed lines.

Vertical forces are again always positive, expelling the magnet. As in the previous case there is a point of equilibrium along the z-axis at z=12.0 mm. As the point of equilibrium it is above the Meissner limit the torus is in a perfect Meissner superconducting state.

Two regions can be observed for the lateral force $F_x$: the outer region in which the magnet is pushed away from the torus and the inner region where it is driven towards the z-axis. A discontinuity can be observed as a white line between both regions. This is also the case for the torque $M_y$.

## V. MAGNET FREE TO ROTATE

Finally, the model has been used to map the angle of equilibrium of the magnet when is located at any position and can freely rotate on the zx-plane. This is shown in Figure 4. It is remarkable that the angle of equilibrium changes sharply from 0º to 90º along the z-axis between z=11.0 and z=11.5 mm. This is the flip effect described by the authors in previous papers[9,10].

Figure 5 shows the force exerted on the permanent magnet when it is oriented at the angle of equilibrium. It can be seen how the inner region constitutes a stable region, with the force directed towards the center of the torus. This not only means that the two equilibrium positions at the z-axis found in the two previous sections are stable, but also any kind of orbit around them, provided that the magnet is able to rotate according to the angle of equilibrium.

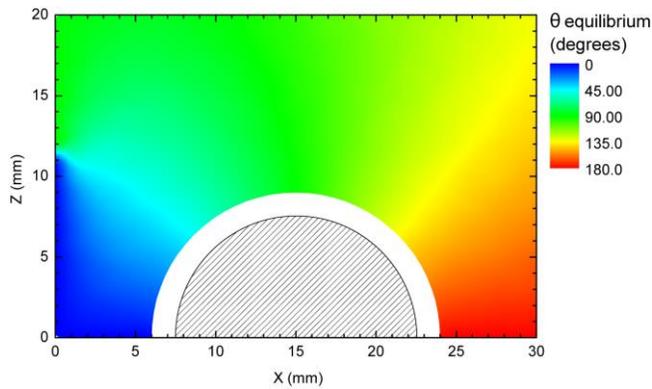

**Figure 4 – Angle of equilibrium for a magnet above a superconducting torus.**

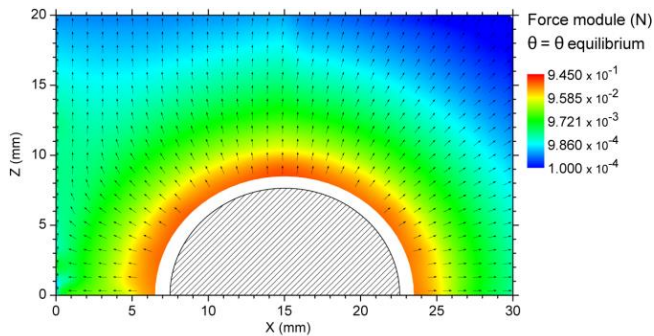

**Figure 5 – Force for a magnet (at the angle of equilibrium) above a superconducting torus. Module is shown in a color scale while direction is shown by arrows.**

An extrapolation of this result suggests the use of a superconducting torus to produce stable orbits for magnetic particles. This includes iron, cobalt or other metallic aggregates.

## VI. CONCLUSIONS

The mapping of the force and torque exerted by a superconducting torus on a permanent magnet shows that there are stable points and orbits where the magnet can levitate. The torus remains superconducting in the Meissner state and not in the mixed state, and therefore the levitation is non-hysteretic.

There are two different points of stability in the axis of symmetry of the torus: one corresponds to a vertically-oriented magnet while the other corresponds to one that is horizontally-oriented. Besides these points of stability in the cases in which the orientation of the magnet is constrained, there are also natural points of stability when the magnet is allowed to rotate freely.

The fact that the stable point of equilibrium for a horizontally-oriented magnet is above that of the vertically-oriented magnet demonstrates that the first case is a better choice for the design of a thrust bearing or a gimbal and the second case is better for journal bearing

**ACKNOWLEDGMENTS**

The research leading to these results has received funding from the European Community's Seventh Framework Programme (FP7/2007-2013) under grant agreement no. 263014.

[1] M.K. Alqadi, F.Y. Alzoubi, H.M. Al-Khateeb, and N.Y. Ayoub, Physica B **404**, 1781-1784 (2009).

[2] F.Y. Alzoubi, M.K. Alqadi, H.M. Al-Khateeb, and N.Y. Ayoub, IEEE Transactions on Applied Superconductivity **17**, 3814-3818 (2007).

[3] D. Palaniappan, Journal of Superconductivity and Novel Magnetism **22**, 471-477 (2009).

[4] E. Diez-Jimenez, J.L. Perez-Diaz, and J.C. Garcia-Prada, Journal of Applied Physics **109**, 063901 (2011).

[5] J.L. Perez-Diaz and J.C. Garcia-Prada, Physica C **467**, 141-144 (2007).

[6] J.L. Perez-Diaz and J.C. Garcia-Prada, Applied Physics Letters **91**, 142503 (2007).

[7] I. Valiente-Blanco, E. Diez-Jimenez, and J.L. Perez-Diaz, Journal of Applied Physics **109**, 07E704 (2011).

[8] E. Diez-Jimenez and J.L. Perez-Diaz, Physica C **471**, 8-11 (2011).

[9] E. Diez-Jimenez, B. Sander, L. Timm, and J.L. Perez-Diaz, Physica C **471**, 229-232 (2010).

[10] J.L. Perez-Diaz, J.C. Garcia-Prada, and J.A. Diaz, Physica C **469**, 252-255 (2009).

[11] E. Diez-Jimenez, J.L. Perez-Diaz, and J.C. Garcia-Prada, IEEE Transactions on Applied Superconductivity, **22**, 9003106, (2012).